\documentclass[11pt]{article}

\usepackage{amsmath} 
\usepackage{amsthm} 
\usepackage{amssymb}	
\usepackage{graphicx} 
\usepackage{multicol} 
\usepackage{multirow}
\usepackage{color}
\usepackage[dvips,letterpaper,margin=1in,bottom=1in]{geometry}

\usepackage[utf8]{inputenc}
\usepackage[english]{babel}

\usepackage{diagbox}
\usepackage{mathtools}

\newtheorem{theorem}{Theorem}[section]

\newtheorem{claim}[theorem]{Claim}
\newtheorem{definition}{Definition}[section]

\newtheorem{remark}{Remark}[section]
\newtheorem{problem}{Problem}

\newcommand{\braket}[2]{\left< #1 \vphantom{#2} \middle| #2 \vphantom{#1} \right>} 

\newcommand{\abs}[1]{\left| #1 \right|}
\newcommand{\Abs}[1]{\left\lVert #1 \right\rVert}
\newcommand{\ket}[1]{\left| #1 \right>}
\newcommand{\bra}[1]{\left< #1 \right|}

\newcommand{\tr} {\operatorname{tr}}
\newcommand{\poly} {\operatorname{poly}}
\newcommand{\polylog} {\operatorname{polylog}}
\newcommand{\rank} {\operatorname{rank}}

\usepackage{latexsym}
\usepackage{CJK}

\usepackage{enumerate}

\usepackage{algorithm}
\usepackage{algpseudocode}
\renewcommand{\algorithmicrequire}{ \textbf{Input:}} 
\renewcommand{\algorithmicensure}{ \textbf{Output:}} 

\usepackage{stmaryrd}
\usepackage{booktabs}

\usepackage{hyperref}
\newcommand{\footremember}[2]{%
    \footnote{#2}
    \newcounter{#1}
    \setcounter{#1}{\value{footnote}}%
}

\usepackage{cite}
\usepackage{bbm}

\usepackage{tablefootnote}
\usepackage{threeparttable}
\usepackage{scrextend}

\usepackage{tikz}
\usetikzlibrary{positioning}

\begin{document}

    \title{Quantum Algorithm for Fidelity Estimation}
        \author{
            Qisheng Wang
            \footremember{1}{Qisheng Wang is with the Department of Computer Science and Technology, Tsinghua University, Beijing, China (e-mail: \url{QishengWang1994@gmail.com}).}
            \and Zhicheng Zhang
            \footremember{2}{Zhicheng Zhang is with Centre for Quantum Software and Information, University of Technology Sydney, Australia (e-mail: \url{iszczhang@gmail.com}). Part of this work was done while the author was at the University of Chinese Academy of Sciences, China.}
            \and Kean Chen
            \footremember{3}{Kean Chen is with the State Key Laboratory of Computer Science, Institute of Software, Chinese Academy of Sciences, China and the University of Chinese Academy of Sciences, China (e-mail: \url{chenka@ios.ac.cn}).}
            \and Ji Guan
            \footremember{4}{Ji Guan is with the State Key Laboratory of Computer Science, Institute of Software, Chinese Academy of Sciences, China (e-mail: \url{guanj@ios.ac.cn}).}
            \and Wang Fang
            \footremember{5}{Wang Fang is with the State Key Laboratory of Computer Science, Institute of Software, Chinese Academy of Sciences, China and the University of Chinese Academy of Sciences, China (e-mail: \url{fangw@ios.ac.cn}).}
            \and Junyi Liu
            \footremember{6}{Junyi Liu is with the State Key Laboratory of Computer Science, Institute of Software, Chinese Academy of Sciences, China and the University of Chinese Academy of Sciences, China (e-mail: \url{liujy@ios.ac.cn}).}
            \and Mingsheng Ying \footremember{7}{Mingsheng Ying is with the State Key Laboratory of Computer Science, Institute of Software, Chinese Academy of Sciences, Beijing, China, and also with the Department of Computer Science and Technology, Tsinghua University, Beijing, China (e-mail: \url{yingms@ios.ac.cn}).}
        }

        \date{}
        \maketitle

    \begin{abstract}
    For two unknown mixed quantum states $\rho$ and $\sigma$ in an $N$-dimensional Hilbert space, computing their 
fidelity $F(\rho,\sigma)$ is a basic problem with many important applications in quantum computing and quantum information, for example verification and characterization of the outputs of a quantum computer, and design and analysis of quantum algorithms.
In this paper, we propose a quantum algorithm that solves this problem in $\poly(\log (N), r, 1/\varepsilon)$ time, where $r$ is the lower rank of $\rho$ and $\sigma$, and $\varepsilon$ is the desired precision, provided that the purifications of $\rho$ and $\sigma$ are prepared by quantum oracles. This algorithm exhibits an exponential speedup over the best known algorithm (based on quantum state tomography) which has time complexity polynomial in $N$.
    \end{abstract}

    \textbf{Keywords: quantum computing, quantum algorithms, quantum fidelity, quantum states.}

    \newpage

    \tableofcontents
    \newpage

\section{Introduction}
{Q}{uantum} computers are believed to have more computing power than classical machines as quantum algorithms have been proven to achieve significant speedups over the best known classical algorithms for solving certain problems. However, only a few of them reach exponential speedups, such as the celebrated Shor's algorithm for integer factorization \cite{Sho94}, the HHL algorithm for solving systems of linear equations~\cite{HHL09} and those for quantum simulation~\cite{Fey82, AL97, FKW02, KJL08}. This paper proposes a quantum algorithm to efficiently estimate the quantum state fidelity on a 
quantum computer. Compared to classical and even known quantum algorithms for the same task, the algorithm can be exponentially faster.

Estimating the quantum state fidelity is a basic problem in quantum computing and quantum information,
as this quantity is one of the most popular and important measures of the ``closeness" of two unknown quantum states~\cite{Uhl76, Joz94, NC10}. Formally, the fidelity of two mixed quantum states $\rho$ and $\sigma$ is defined as
\[
F(\rho, \sigma) = \tr\left(\sqrt{\sqrt{\sigma}\rho\sqrt{\sigma}}\right).
\]
Quantum state fidelity has been applied in many different  fields, such as quantum information processing \cite{TV11}, quantum engineering \cite{RCR21} and  quantum  machine learning \cite{SR21}.
It also plays a necessary and essential role in 
verifying and characterizing the output state of a quantum computer.

We propose an efficient quantum algorithm for fidelity estimation of quantum states, stated as follows.

\begin{theorem}[Informal]
    There is a quantum algorithm that, given ``purified quantum query access'' to two $N$-dimensional quantum states $\rho$ and $\sigma$, computes their fidelity within additive error $\varepsilon$ with time complexity $\poly(\log(N), r, 1/\varepsilon)$, where $r$ is the lower rank of $\rho$ and $\sigma$.
\end{theorem}

The ``purified quantum query access'' model is widely used in quantum computational complexity theory and quantum algorithms (e.g., \cite{Wat02,BKL+19,vAG19,GL20,GLM+20,ARSW21,GHS21,SH21}), where mixed quantum states (density operators) are given by quantum circuits (oracles) that prepare their purifications (see Section \ref{sec:purified-quantum-access-query} for a formal definition).
From the perspective of computational complexity theory, the purified quantum query access model is useful when comparing quantum and classical algorithms, especially in defining complexity classes (see \cite{Wat08} for example).
This model was introduced in \cite{GL20} to the context of density operator testing. The basic idea behind the model is that a density operator can be understood as the output of certain quantum process. If we are able to simulate this process on a quantum computer, then it indeed provides purified quantum query access to the density operator.

In this Introduction, we will first review the existing approaches for fidelity estimation and discuss about its computational hardness.
Then an outline of our quantum algorithm for fidelity estimation will be given in Section \ref{sec:overview-algorithm}. The details of the algorithm will be carefully described in the subsequent sections.

\subsection{Existing Approaches for Fidelity Estimation}

There are no known efficient methods for estimating the fidelity $F(\rho, \sigma)$ in the general case.
A straightforward way is to first obtain a complete classical description of quantum states as density matrices by quantum state tomography \cite{DM97, dAMP03}, and then calculate the fidelity by matrix arithmetic operations. However, this kind of approach requires resources increasing exponentially with the scale of the quantum system, even if quantum states are restricted to be low-rank. A slightly more efficient tomography can be applied for low-rank quantum states \cite{GLF10,OW16,HHJ+17}.
Several approaches for estimating  $F(\rho,\sigma)$ have been proposed for the special case where $\rho$ or $\sigma$ is pure. The first one is the SWAP test \cite{BCW01, KHC19} using the Hadamard and Toffoli gates, which computes the value of $\tr(\rho\sigma)$. Then a more practical technique called entanglement witnesses \cite{TYKI06, GLGP07, GT09} was introduced to compute the fidelity with few measurements, but only works for some specific pure quantum states. This limitation was overcome in ~\cite{FL11, dSLP11} where a direct fidelity estimation method was developed for arbitrary pure quantum states.

Recently, suspecting that computing fidelity of quantum states in the general case could be hard, several variational quantum algorithms for fidelity estimation have been  successively proposed \cite{CPCC20,CSZW22,TV21}. Naturally, the efficiency of these algorithms is unknown as training variational quantum algorithms is known to be \textbf{NP}-hard \cite{BK21}.

\subsection{Hardness of Fidelity Estimation}

Computing $F(\rho, \sigma)$ in general is known to be \textbf{QSZK}-hard \cite{Wat02}, where \textbf{QSZK} (Quantum Statistical Zero-Knowledge) is a complexity class which contains \textbf{BQP}.
   A restricted version called \textsc{Low-Rank Fidelity Estimation}, namely estimating the fidelity of low-rank quantum states, was shown in \cite{CPCC20} to be \textbf{DQC1}-hard. Here, \textbf{DQC1} (Deterministic Quantum Computing with 1 Clean Bit) is the complexity class of problems that can be efficiently solved in the one-clean-qubit model of quantum computing~\cite{KL98}, which is commonly believed to be strictly contained in \textbf{BQP}. However, it was observed in ~\cite{Mor17, FKM18} that efficient classical simulation of \textbf{DQC1}-complete problems implies that the polynomial hierarchy collapses to the second level, which is not believed to be true.
   Our algorithm presented in this paper shows that \textsc{Low-Rank Fidelity Estimation} can be efficiently solved by quantum computers with complexity logarithmic in the dimension of quantum states, and therefore establishes a \textbf{BQP} upper bound for it. We summarize the results about for the hardness of fidelity estimation in Table \ref{tab:hard}.

\begin{table}[htp]
\caption{Hardness of quantum fidelity estimation. }
\label{tab:hard}
\centering
\begin{tabular}{ c c c }
\hline
Restriction       & Lower Bound        & Upper Bound          \\ \hline
General Case  & \textbf{QSZK}-hard \cite{Wat02} & \textbf{EXPTIME} \tablefootnote{By direct classical simulation of quantum computation and matrix arithmetic operations.}     \\ \hline
Low-Rank Case & \textbf{DQC1}-hard \cite{CPCC20} & \textbf{BQP} (this paper)         \\ \hline
\end{tabular}
\end{table}

\subsection{Overview of Our Algorithm} \label{sec:overview-algorithm}

The input model used in our quantum algorithm is usually called the ``purified quantum query access'' model, which provides (mixed) quantum states by quantum oracles that prepare their purifications. This conventional model is commonly used in quantum algorithms, e.g., \cite{BKL+19,vAG19,GL20,GLM+20,ARSW21,GHS21,SH21}, and will be introduced in Section \ref{sec:purified-quantum-access-query}.

We sketch the basic idea of our algorithm here, and then describe it in detail later. The goal is to obtain the subnormalized quantum state $\sqrt{\sqrt\sigma\rho\sqrt\sigma}$ with a certain  probability, and then we can estimate the  fidelity $F(\rho,\sigma)$ by evaluating its trace. To achieve this, our algorithm consists of two parts. The first part is a technique that prepares a quantum state that block-encodes the square root of a positive semidefinite operator block-encoded
in another quantum state.
Here, the notion of block-encoding (see Definition~\ref{def:block-encoding}) is allowed for  encoding a general matrix into a scaled block in a larger matrix, and thus extends the block-encoding defined for unitary operators in~\cite{LC19,GSLW19}.
Specifically, suppose that we are given a quantum state which is a block-encoding of a positive semidefinite operator $A$. Then we are able to prepare another quantum state which is a block-encoding of $\sqrt{A}$.
This is achieved through the technique of quantum singular value transformation \cite{GSLW19} combined with our new idea of constructing density operators (instead of unitary operators) as block-encodings.
The second part uses the technique developed in the first part for multiple times. Suppose we are given two $N$-dimensional quantum states $\rho$ and $\sigma$. Since $\sigma$ is a block-encoding of itself, we can prepare a quantum state which is a block-encoding of $\sqrt{\sigma}$. This also gives a unitary operator $V_\sigma$ that is a block-encoding of $\sqrt{\sigma}$ by Lemma 25 of~\cite{GSLW19}. After applying $V_\sigma$ on $\rho$, we obtain a quantum state which is a block-encoding of $\sqrt{\sigma}\rho\sqrt{\sigma}$. Applying the technique again, we obtain a quantum state which is a block-encoding of $\sqrt{\sqrt{\sigma}\rho\sqrt{\sigma}}$. Finally, we can compute the fidelity $F(\rho, \sigma) = \tr\left(\sqrt{\sqrt{\sigma}\rho\sqrt{\sigma}}\right)$ by quantum amplitude estimation~\cite{BHMT02}.

An important factor in the performance of our fidelity estimation algorithm is the rank of quantum states. Let $r$ be the lower rank of $\rho$ and $\sigma$.
Our algorithm has a time complexity $\poly(\log (N), r, 1/\varepsilon)$, where $\varepsilon$ is the additive error. As the rank $r$ grows, errors become harder to deal with and the algorithm becomes less efficient. Our algorithm exponentially outperforms the known classical and even quantum algorithms in the case that one of the two quantum states is low-rank, say $r = \polylog(N)$. A quantum state is said to be low-rank if it is supported on a low-dimensional space. Low-rank quantum states are nearly pure and with low entropy, which appear in various important  physical settings~\cite{GLF10, LH08}.

Compared to those approaches based on quantum state tomography \cite{DM97, dAMP03, GLF10}, the strength of our algorithm is that it only uses $\polylog(N)$ qubits, and does not involve classical matrix representations of quantum states. On the other hand, our algorithm works in a more general case than the SWAP test~\cite{BCW01, KHC19}, entanglement witnesses \cite{TYKI06, GLGP07, GT09} and the direct fidelity estimation \cite{FL11, dSLP11} where one of the states is required to be pure in the latter case. 
A comparison of various fidelity estimation algorithms is presented in Table \ref{tab:cmp}, where $N$ is the dimension of quantum states and $r$ is the lower rank of them. Note that $r = 1$ means one of the quantum states is pure.
It should be pointed out that our quantum algorithm requires purified quantum query access to quantum states, while most of other quantum algorithms in Table \ref{tab:cmp} require identical copies of quantum states. We argue that our quantum algorithm is particularly useful when the quantum processes that produce these quantum states can be implemented on a quantum computer. Furthermore, given purified quantum query access to density operators, the algorithms that require identical copies can be easily converted to ones with purified quantum query access, while preserving their computational complexity. In this sense, our algorithm can be exponentially faster than the others listed in Table \ref{tab:cmp}. Also,  if we adopt the conventional input model commonly used in quantum computational complexity theory
where mixed quantum states are given by classical descriptions of quantum circuits that produce their purifications (see \cite{Wat08} for example), then our quantum algorithm can be regarded as a candidate of quantum advantages over classical computing.

\begin{table*}
\caption{\label{tab:cmp}Comparison of algorithms for quantum fidelity estimation. }
\centering
\scriptsize
\begin{tabular}{ c c c c }
Algorithm                              & Complexity      & Prerequisites                       & Required Operations                \\ \hline
Quantum State Tomography \cite{DM97, dAMP03, GLF10}               &  $\poly(N)$         & None                    & Pauli Measurements           \\ \hline
Low-rank Quantum State Tomography \cite{GLF10}      &  $\poly(N, r)$      & $r$ is small                         & Pauli Measurements           \\ \hline
SWAP Test \cite{BCW01, KHC19}                              & $\polylog(N)$      & $r = 1$                              & Arbitrary Quantum Operations \\ \hline
Entanglement Witnesses \cite{TYKI06, GLGP07, GT09}                 & $\polylog(N)$      & Specific Pure States & Pauli Measurements           \\ \hline
Direct Fidelity Estimation \cite{FL11, dSLP11}             & $\poly(N)$         & $r = 1$                              & Pauli Measurements           \\ \hline
Variational Quantum Algorithm \cite{CPCC20, CSZW22, TV21} & N/A             & $r$ is small                         & Arbitrary Quantum Operations \\ \hline
Our Algorithm                           & $\poly(\log(N), r)$ & $r$ is small                         & Arbitrary Quantum Operations \\
\end{tabular}
\end{table*}

\subsection{Related Work}

Estimating the ``closeness'' of quantum states is an important problem in quantum information; in particular, it is closely related to quantum property testing \cite{MdW16}. Except those mentioned above, here we briefly discuss some other related work. A method to test the closeness of $N$-dimensional mixed quantum states with respect to the trace distance and fidelity was proposed in \cite{BOW19}, using $O(N/\varepsilon^2)$ and $O(N/\varepsilon)$ copies of quantum states, respectively. (It is worth mentioning that testing the closeness is weaker than estimating the closeness: taking the fidelity for example, the former task is to distinguish whether the fidelity is close to $1$, while the latter one is to find the approximate value of the fidelity.) On the other hand, a quantum algorithm for estimating the trace distance between two quantum states was developed in \cite{GL20} using $O(N/\varepsilon)$ queries to quantum oracles in the ``purified quantum query access'' model.

Testing other quantum properties is also a widely studied topic. Most of them focus on the sample complexity (i.e., the number of copies of quantum states that are used in the testing).
For example, quantum state tomography and its extensions have been studied in \cite{OW15,OW16,HHJ+17,OW17,BOW19};
 the sample complexity of testing the orthogonality of two pure quantum states was investigated in \cite{KLL+17};
and the sample complexity of estimating the von Neumann entropy and the quantum R\'{e}nyi entropy was examined in \cite{AISW19}.
Although these results were obtained employing the multiple-copy input model, several  others were established in the ``purified quantum query access'' model; for instance, the quantum query complexity of estimating the von Neumann entropy was considered in \cite{GL20,GHS21}, and a quantum query algorithm for estimating the quantum R\'{e}nyi entropy was proposed in \cite{SH21}.

In addition, quantum algorithms for testing classical statistical properties have been extensively studied. The first  quantum algorithms for testing closeness and identity of probability distributions was given in \cite{BHH11}, which were then improved by \cite{Mon15} and \cite{CFMdW10}. Quantum approach for estimating classical entropies was  systematically studied in  \cite{LW19}.

\subsection{Recent Developments}
After the work described in this paper, several improvements have been made, compared to the complexity $\tilde O(r^{12.5}/\varepsilon^{13.5})$\footnote{$\tilde O(\cdot)$ suppresses polylogarithmic factors.} stated in Theorem \ref{thm:main}.
\begin{itemize}
    \item Wang et al. \cite{WGL+22} improved the quantum query complexity of fidelity estimation in the purified quantum query access model to $\tilde O(r^{6.5}/\varepsilon^{7.5})$. Moreover, they proposed quantum algorithms for estimating a wide range of quantum entropies and distances.
    \item In the concurrent work of Gily\'{e}n and Poremba \cite{GP22}, they improved the quantum query complexity of fidelity estimation to $\tilde O(r^{2.5}/\varepsilon^5)$. Moreover, they converted their quantum algorithm for fidelity estimation in the purified quantum access model to the one using $\tilde O(r^{5.5}/\varepsilon^{12})$ identical copies of quantum states based on the technique of density matrix exponentiation \cite{LMR14,KLL+17}.
\end{itemize}

\subsection{Organization of This Paper}

The rest of this paper is organised as follows. Section \ref{sec:preliminaries}  will provide necessary preliminaries. We will formally state our main result in Section \ref{sec:main-result}. Then, we will show our technique for solving  square roots of positive semidefinite operators in Section \ref{sec:sqrt}. Our quantum algorithm for fidelity estimation and an analysis of its complexity will be elaborated in Section \ref{sec:fidelity-estimation}. In Section \ref{sec:hardness}, we will further discuss the hardness of quantum fidelity estimation. We will conclude in Section \ref{sec:conclusion} with a brief discussion about applications and extensions of our algorithm.

\section{Preliminaries} \label{sec:preliminaries}

\subsection{Block-encoding and its extension}

The notion of block-encoding was defined in \cite{LC19,GSLW19} to describe quantum unitary operators, in which certain operators of interest are encoded as scaled matrix blocks.
It is proved to be a useful tool in developing quantum algorithms. Different from the original form of block-encoding, our quantum algorithm needs to block-encode a positive semidefinite operator in a density operator rather than in a unitary operator.
For this purpose, we extend the definition of block-encodings to  general operators as follows.

\begin{definition} [Block-encoding] \label{def:block-encoding}
    Suppose $A$ is an $n$-qubit operator, $\alpha, \varepsilon \geq 0$ and $a \in \mathbb{N}$. An $(n+a)$-qubit operator $B$ is said to be an $(\alpha, a, \varepsilon)$-block-encoding of $A$, if
    \[
        \Abs{ \alpha \prescript{}{a}{\bra 0} B \ket 0_a - A } \leq \varepsilon,
    \]
    where the operator norm is defined by $$\Abs{A} = \sup_{\sqrt{\braket{\psi}{\psi}} = 1} \sqrt{\bra{\psi} A^\dagger A\ket{\psi}}$$
and $A^\dagger$ is the Hermitian conjugate of $A$.
    \end{definition}

\begin{remark}
Note that the above definition of block-encodings coincides with that in \cite{LC19,GSLW19}, whenever the block-encoding $B$ is restricted to be unitary. In our extended definition, however, we allow the block-encoding $B$ to be a density operator. When an operator $A$ is block-encoded in a density operator $\rho$, it means that $A$ can be obtained by measuring $\rho$ on a subsystem (subscripted by $a$ in Definition~\ref{def:block-encoding}) and post-selecting the outcome $0$.
\end{remark}

We will use this  extended definition of block-encodings throughout the paper when describing both unitary and density operators.

\subsection{Purified quantum query access} \label{sec:purified-quantum-access-query}

The ``purified quantum query access'' model is widely used in quantum computational complexity and quantum algorithms (e.g., \cite{Wat02,BKL+19,vAG19,GL20,GLM+20,ARSW21,GHS21,SH21}). In this model, mixed quantum states are given by quantum circuits (oracles) that prepare their purifications. Formally, suppose $\rho$ is a mixed quantum state in an $N$-dimensional Hilbert space. A quantum unitary oracle $O_\rho$ that prepares $\ket{\rho}$ is given as
\[
    \ket{\rho} = O_\rho \ket{0}_n \ket{0}_{n_\rho},
\]
where $N = 2^n$, and $\ket{0}_{n_\rho}$ are ancilla qubits. Here, we write $\ket{0}_a$ to denote $\ket{0}^{\otimes a}$, with the subscript $a$ indicating which (and how many) qubits are involved in the Dirac symbol.
This notation is convenient in analysis when more than two disjoint sets of qubits are considered simultaneously. Then $\rho$ is obtained from its purification by tracing out those ancilla qubits:
\[\rho = \tr_{n_\rho}(\ket\rho\bra\rho).\]
We assume that $n_\rho$ is a polynomial in $n$~\footnote{Theoretically, any $n$-qubit mixed quantum state has a purification with at most $n$ ancilla qubits, so it is sufficient to assume that the number of ancilla qubits is no more than $n$. However, it could be more convenient to use more than $n$ ancilla qubits in order to prepare a purification of an $n$-qubit mixed quantum state in practice. That is why we make a more relaxed assumption on the number of ancilla qubits, which is just a polynomial in $n$.}.
In the following, a unitary operator $U$ is said to prepare a mixed quantum state, if $U$ prepares its purification.

Next, we will introduce a useful tool, which can convert a unitary operator that prepares a density operator to another unitary operator that is a block-encoding of the density operator \cite{LC19,vAG19,GSLW19}.

\begin{theorem} [Block-encoding of density operators, Lemma 25 of \cite{GSLW19}] \label{thm:block-encoding-density-operator}
    Suppose $U$ is an $(n+a)$-qubit unitary operator that prepares an $n$-qubit density operator $\rho$. Then there is a $(2n+a)$-qubit unitary operator $\tilde U$, which is a $(1, n+a, 0)$-block-encoding of $\rho$, using $1$ query to each of $U$ and $U^\dag$ and $O(a)$ elementary quantum gates.
\end{theorem}

\begin{remark}
It is worth mentioning that there is no known method that conversely converts a unitary operator, which is a block-encoding of a density operator, to another unitary operator, which prepares the  density operator. This motivates us to directly manipulate density operators rather than unitary operators. It is also why we chose to extend the definition of block-encodings (see Definition \ref{def:block-encoding}).
\end{remark}

\subsection{Polynomial eigenvalue transformation}

Quantum singular value transformation (QSVT) \cite{GSLW19} is a powerful framework in designing quantum algorithms. What we need is the QSVT technique for polynomial eigenvalue transformation.

\begin{theorem} [Polynomial eigenvalue transformation, Theorem 31 of \cite{GSLW19}] \label{thm:polynomial-eigenvalue-transformation}
Suppose $U$ is a unitary operator, which is an $(\alpha, a, \varepsilon)$-block-encoding of Hermitian operator $A$. If $\delta \geq 0$ and $P \in \mathbb{R}[x]$ is a polynomial of degree $d$ such that $\abs{P(x)} \leq 1/2$ for all $x \in [-1, 1]$. Then there is a quantum circuit $\tilde U$, which is a $(1, a+2, 4d\sqrt{\varepsilon/\alpha}+ \delta)$-block-encoding of $P(A/\alpha)$, using $d$ queries to $U$ and $U^\dag$, $1$ query to controlled-$U$, and $O((a+1)d)$ elementary quantum gates. Moreover, the description of $\tilde U$ can be computed by a classical Turing machine in $\poly(d, \log(1/\delta))$ time.

Especially, if $P$ is an even or odd polynomial, then $\abs{P(x)} \leq 1/2$ for $x \in [-1, 1]$ can be relaxed to $\abs{P(x)} \leq 1$ for $x \in [-1, 1]$.
\end{theorem}

In order to apply the polynomial eigenvalue transformation for our purpose, the polynomial approximation of negative power functions is required.

\begin{theorem} [Polynomial approximation of negative power functions, Corollary 67 in the full version of \cite{GSLW19}] \label{thm:polynomial-approx}
Let $\delta, \varepsilon \in (0, 1/2]$, $c > 0$ and $f(x) = \delta^c x^{-c} / 2$. Then there is an even/odd polynomial $P \in \mathbb{R}[x]$ of degree $O\left(\frac{\max\{1, c\}}{\delta} \log\left(\frac{1}{\varepsilon}\right)\right)$ such that $\abs{P(x)-f(x)} \leq \varepsilon$ for $x \in [\delta, 1]$ and $\abs{P(x)} \leq 1$ for $x \in [-1, 1]$.
\end{theorem}

\section{Main Result} \label{sec:main-result}
As discussed in  Section \ref{sec:purified-quantum-access-query}), we work in the ``purified quantum query access'' model. Suppose two (mixed) quantum states $\rho$ and $\sigma$ in an $N$-dimensional Hilbert space are given by their  corresponding purifications $\ket{\rho}$ and $\ket{\sigma}$; that is, two quantum unitary oracles (circuits) $O_\rho$ and $O_\sigma$ that prepare $\ket{\rho}$ and $\ket{\sigma}$, respectively, are assumed as follows:
\[
    \ket{\rho}=O_{\rho}\ket 0_n \ket 0_{n_\rho}, \qquad \ket{\sigma}=O_{\sigma}\ket 0_n \ket 0_{n_\sigma},
\]
where $N = 2^n$,  $\ket 0_{n_\rho}$ and $\ket 0_{n_\sigma}$ are ancilla qubits, and $n_\rho$ and $n_\sigma$ are polynomials in $n$.
Then $\rho$ and $\sigma$ are obtained from their purifications by tracing out their corresponding ancilla qubits:
\[\rho = \tr_{n_\rho}(\ket\rho\bra\rho), \qquad \sigma = \tr_{n_\sigma}(\ket\sigma\bra\sigma). \]

In this paper, we use the following definition for $\tilde O$:
\[
\tilde O_{d, e}(f(a, b, c)) = O(f(a, b, c)  \polylog(f(a, b, c), d, e)).
\]
Then our main result can be stated as the following:

\begin{theorem} \label{thm:main}
    Given quantum oracles $O_\rho$ and $O_\sigma$ that prepare $N$-dimensional quantum states $\rho$ and $\sigma$, respectively, there is a quantum algorithm that computes the fidelity $F(\rho, \sigma)$ within additive error $\varepsilon$ using $\tilde O_{r,1/\varepsilon}\left(r^{12.5}/\varepsilon^{13.5}\right)$ queries to these oracles and $\tilde O_{N,r,1/\varepsilon}\left(r^{12.5}/\varepsilon^{13.5}\right)$ additional elementary quantum gates, where $r$ is the lower rank of $\rho$ and $\sigma$.
\end{theorem}

Our algorithm will be presented in the following way. First, we develop a technique to compute the square root of a positive semidefinite operator (see Section \ref{sec:sqrt}). Then, we apply this  technique multiple times in order to estimate the fidelity (see Section \ref{sec:fidelity-estimation}).

\section{Square Root of Positive Semidefinite Operators} \label{sec:sqrt}

The key technique of our algorithm is to compute the square root of a positive semidefinite operator stored in a quantum state in the sense of block-encoding.
 This technique can be described as the following:

\begin{theorem}[Square root of positive semidefinite operators block-encoded in density operators] \label{thm:sqrt}
Suppose:
\begin{enumerate}
  \item $\rho$ is an $(n+a)$-qubit density operator with an $(n+a+b)$-qubit pure quantum state $\ket{\rho}$ being its purification; that is, $\rho = \tr_b(\ket \rho \bra \rho)$. An $(n+a+b)$-qubit unitary operator $U_\rho$ is given to prepare $\ket \rho = U_\rho \ket 0$;
  \item $A$ is an $n$-qubit positive semidefinite operator such that $\rho$ is a $(1, a, 0)$-block-encoding of $A$.
\end{enumerate}
Then for every real number $\delta, \varepsilon \in (0, 1/2]$, there is an $O(n+a+b)$-qubit quantum circuit $U_\varrho$ such that
\begin{itemize}
  \item $U_\varrho$ uses $O(d)$ queries to (controlled-)$U_\rho$ and its inverse and $O(d(n+a+b))$ elementary quantum gates, where $d = O(\log(1/\varepsilon)/\delta)$; and
  \item $U_\varrho$ prepares the purification $\ket\varrho = U_\varrho \ket 0$ of a $(4\delta^{-1/2}, O(n+a+b), \Theta(\delta^{1/2}+\varepsilon \delta^{-1/2}))$-block-encoding $\varrho$ of $\sqrt{A}$.
  \item The description of $U_\varrho$ can be computed by a classical Turing machine in $\poly(d)$ time.
\end{itemize}
\end{theorem}

\begin{proof}
Recall Theorem \ref{thm:block-encoding-density-operator} that a unitary operator that prepares a mixed quantum state $\rho$ implies another unitary operator that is a block-encoding of $\rho$. Then there is a unitary operator $\tilde U_A$, which is a $(1, n+a+b, 0)$-block-encoding of $\rho$, and therefore a $(1, n+2a+b, 0)$-block-encoding of $A$, using $1$ query to $U_\rho$, and $O(a+b)$ elementary quantum gates.

Let $f(x) = (\delta/x)^{1/4}/2$.
By Theorem \ref{thm:polynomial-approx}, there is an even polynomial $P(x)$ of degree $O(d)$ such that $\abs{P(x)-f(x)} \leq \varepsilon$ for every $x \in [\delta, 1]$ and $\abs{P(x)} \leq 1$ for every $x \in [-1, 1]$. By Theorem \ref{thm:polynomial-eigenvalue-transformation}, there is a unitary operator $\tilde U_{P(A)}$, which is a $(1, O(n+a+b), \varepsilon)$-block-encoding of $P(A)$ using $d$ queries to $\tilde U_A$ and $O(d(n+a+b))$ elementary quantum gates.

Applying $\tilde U_{P(A)}$ on $\rho$, we obtain a density operator $\varrho$, which is a $(1, O(n+a+b), 0)$-block-encoding of $A(P(A))^2$, and therefore a $(4\delta^{-1/2}, O(n+a+b), \Theta(\delta^{1/2} + \varepsilon \delta^{-1/2}))$-block-encoding of $\sqrt{A}$. To see this, we need to show that
\[
\Abs{4\delta^{-1/2}A(P(A))^2 - \sqrt{A}} \leq \Theta(\delta^{1/2} + \varepsilon \delta^{-1/2}).
\]
Since $0 \leq A \leq I$, it is sufficient to show that
\[
\abs{4\delta^{-1/2}x(P(x))^2 - \sqrt{x}} \leq \Theta(\delta^{1/2} + \varepsilon \delta^{-1/2})
\]
for every $x \in [0, 1]$. We consider two cases as follows.

\textbf{Case 1}. $x \in [\delta, 1]$. In this case,
\begin{align*}
    \abs{4\delta^{-1/2} x(P(x))^2 - \sqrt{x}}
    & = \abs{4\delta^{-1/2}x(P(x))^2 -4\delta^{-1/2}x(f(x))^2} \\
    & \leq 4 \delta^{-1/2} \abs{x} \abs{P(x)+f(x)} \abs{P(x)-f(x)} \\
    & \leq 8 \delta^{-1/2} \varepsilon.
\end{align*}

\textbf{Case 2}. $x \in [0, \delta]$. In this case,
\begin{align*}
    \abs{4\delta^{-1/2}x(P(x))^2 - \sqrt{x}} \leq \abs{4\delta^{-1/2}x(P(x))^2} + \abs{\sqrt{x}}
    \leq 5 \delta^{1/2}.
\end{align*}
These yield the proof.
\end{proof}



Theorem~\ref{thm:sqrt} is derived following  the basic procedure of
quantum singular value transformation (QSVT)~\cite{GSLW19}. But a different idea we used in Theorem~\ref{thm:sqrt} is the extension of the block-encoding for \textit{unitary operators} employed in QSVT to that for \textit{density operators}. This new idea enables us to obtain a better complexity. Specifically,
if we try to derive the $\sqrt{A}$ in
Theorem~\ref{thm:sqrt} by QSVT in a similar way of implementing the power function (in our case, the square root function) of an Hermitian matrix block-encoded in a unitary operator~\cite{GSLW19,CGJ19,Gil19} (which is, for example, later used to implement the Petz recovery channels~\cite{GLM+20}), we will meet an additional restriction of $I/\kappa \leq A \leq I$ for some $\kappa > 0$~\cite{CGJ19}. As a result, an unfavorable factor $\kappa$ is introduced in the complexity, and $\kappa$ can be arbitrarily large for any density operator $A$.
In contrast, Theorem~\ref{thm:sqrt} circumvents this difficulty by preparing density operators as block-encodings rather than unitary operators as in QSVT, and thus makes the complexity of our algorithm independent of the parameter $\kappa$.

Our new idea
brings another benefit --- statistical properties of the operator block-encoded in the density operator of a quantum state can be extracted more easily by measurements;
while the same task seems hard for the operator block-encoded in a unitary operator (as in QSVT).
For example, given that $A$ is block-encoded in a density operator $\varrho$ of a mixed quantum state, whose purification is prepared by a unitary operator $U_\varrho$,
the trace $\tr(A)$ can be simply evaluated by quantum amplitude estimation \cite{BHMT02} (see step 4 in Section \ref{sec:fidelity-estimation} below).
However, computing $\tr(A)$ seems to be hard if $A$ is block-encoded in an $n$-qubit unitary operator $U$, as there is no known efficient quantum algorithm even to compute $\tr(U)$ within additive error $1/\poly(n)$ --- the best known approach has additive error $2^n/\poly(n)$~\cite{KL98}, which is exponentially worse than required.
Furthermore, computing $\tr(U)/2^n$ was shown to be \textbf{DQC1}-complete \cite{KL98}.

\section{Fidelity Estimation} \label{sec:fidelity-estimation}

\subsection{The Algorithm}

Now we are able to describe the main algorithm for fidelity estimation.
Without loss of generality, we assume that the rank of $\rho$ is lower than or equal to that of $\sigma$ and let $r = \rank(\rho)$. Note that in this case the state $\sigma$ only contributes to the fidelity on the support of $\rho$, because $F(\rho, \sigma) = F(\rho, \Pi_\rho \sigma \Pi_\rho)$, where $\Pi_\rho$ is the projector onto the support of $\rho$.

Our algorithm is presented as Algorithm \ref{algo}.
\algrenewcommand\algorithmicrequire{\textbf{Input:}}
\algrenewcommand\algorithmicensure{\textbf{Output:}}
\begin{algorithm}[!htp]
        \caption{Quantum algorithm for fidelity estimation.}
        \label{algo}
        \begin{algorithmic}[1]
        \Require Quantum oracles $O_\rho$ and $O_\sigma$ that prepare mixed quantum states $\rho$ and $\sigma$, respectively, the desired additive error $\varepsilon > 0$, and $r = \rank(\rho)$.
        \Ensure An approximation of $F(\rho, \sigma)$ within additive error $\varepsilon$ with high probability.

        \State $\delta_\sigma \gets \tilde\Theta(\varepsilon^4/r^4)$.
        \State $\varepsilon_\sigma \gets \tilde\Theta(\varepsilon^4/r^4)$.
        \State $\delta_\eta \gets \tilde\Theta(\varepsilon^6/r^6)$.
        \State $\varepsilon_\eta \gets \tilde\Theta(\varepsilon^6/r^6)$.
        \State $M \gets \tilde\Theta(\varepsilon^{2.5}/r^{3.5})$.
        \State $V_\sigma$, a unitary operator using $O(\log(1/\varepsilon_\sigma)/\delta_\sigma)$ queries to $O_\sigma$ (by Theorem \ref{thm:sqrt}), prepares $\sigma'$ such that $\sigma'$ is a $(4\delta_\sigma^{-1/2}, b, \Theta(\delta_\sigma^{1/2}+\varepsilon_\sigma\delta_\sigma^{-1/2}))$-block-encoding of $\sqrt\sigma$, where $b = O(n+n_\sigma)$.
        \State $W_\sigma$, a unitary operator using $O(1)$ queries to $V_\sigma$ (by Theorem \ref{thm:block-encoding-density-operator}), is a $(4\delta_\sigma^{-1/2}, O(n+n_{\sigma}), \Theta(\delta_\sigma^{1/2}+\varepsilon_\sigma \delta_\sigma^{-1/2}))$-block-encoding of $\sqrt{\sigma}$.
        \State $U_\eta \gets (W_\sigma \otimes I_{n_\rho}) (O_\rho \otimes I_a)$ prepares $\eta$ that (by Claim \ref{lemma:eta-approx}) is a $(16\delta_\sigma^{-1}, a, \Theta(\delta_\sigma^{1/2}+\varepsilon_\sigma \delta_\sigma^{-1/2}))$-block-encoding of $\sqrt{\sigma}\rho\sqrt{\sigma}$, where $a = O(n+n_\sigma)$.
        \State $V_\eta$, a unitary operator using $O(\log(1/\varepsilon_\eta)/\delta_\eta)$ queries to $U_\eta$ (by Theorem \ref{thm:sqrt}), prepares $\eta'$ such that $\eta'$ is a $(4\delta_{\eta}^{-1/2}, c, \Theta(\delta_{\eta}^{1/2}+\varepsilon_{\eta} \delta_{\eta}^{-1/2}))$-block-encoding of $\sqrt{\prescript{}{a}{\bra 0} \eta \ket 0_a}$, where $c = O(n+n_\rho+n_\sigma)$.
        \State $\tilde x \gets x \pm \delta$ with high probability, using $O(M)$ queries to $V_\eta$ (by quantum amplitude estimation \cite{BHMT02}), where $x = \tr \left( \prescript{}{c}{\bra 0} \eta' \ket 0_c \right)$ and
        \[
            \delta = 2 \pi \frac{\sqrt{x(1-x)}}{M} + \frac{\pi^2}{M^2}.
        \]
        \State \Return $16 \tilde x / \sqrt{\delta_{\eta} \delta_{\sigma}}$.
        \end{algorithmic}
    \end{algorithm}

For a better understanding, let us explain it in five steps:

\textbf{Step 1}. Note that $\sigma$ is a $(1, 0, 0)$-block-encoding of itself, and $O_\sigma$ prepares its purification $\ket{\sigma}$. By Theorem \ref{thm:sqrt} and introducing two parameters $\delta_\sigma$ and $\varepsilon_\sigma$, we can obtain a unitary $V_\sigma$ using $O(\log(1/{\varepsilon_\sigma})/{\delta_\sigma})$ queries to $O_\sigma$ that prepares the purification $V_\sigma \ket{0} = \ket{\sigma'}$ of $\sigma'$, and $\sigma'$ is a $(4\delta_\sigma^{-1/2}, b, \Theta(\delta_\sigma^{1/2}+\varepsilon_\sigma \delta_\sigma^{-1/2}))$-block-encoding of $\sqrt{\sigma}$, where $b = O(n+n_{\sigma})$.

\textbf{Step 2}. By Theorem \ref{thm:block-encoding-density-operator}, we can construct a unitary operator $W_\sigma$ using $1$ query to $V_\sigma$ that is a $(1, O(n+n_\sigma), 0)$-block-encoding of $\sigma'$, and therefore a $(4\delta_\sigma^{-1/2}, O(n+n_{\sigma}), \Theta(\delta_\sigma^{1/2}+\varepsilon_\sigma \delta_\sigma^{-1/2}))$-block-encoding of $\sqrt{\sigma}$. By applying $W_\sigma$ on $\rho$, we obtain a density operator $\eta$ that is a $(16\delta_\sigma^{-1}, a, \Theta(\delta_\sigma^{1/2}+\varepsilon_\sigma \delta_\sigma^{-1/2}))$-block-encoding of $\sqrt{\sigma}\rho\sqrt{\sigma}$, where $a = O(n+n_\sigma)$. In other words, $U_\eta = (W_\sigma \otimes I_{n_\rho})(O_\rho \otimes I_a)$ can prepare the purification $U_\eta\ket{0} = \ket{\eta}$ of $\eta$. To see this, we note that $\eta$ is a $(1, a, 0)$-block-encoding of $\prescript{}{a}{\bra 0} \eta \ket 0_a = \sigma'_b \rho \left(\sigma'_b\right)^\dag$, where $\sigma'_b = \prescript{}{b}{\bra 0} \sigma' \ket 0_b$. Here, we note that $\sigma'_b \rho \left(\sigma'_b\right)^\dag$ is a scaled approximation of $\sqrt{\sigma} \rho \sqrt{\sigma}$ (see Claim \ref{lemma:eta-approx} for details).

\textbf{Step 3}. Similar to the previous, by Theorem \ref{thm:sqrt} and introducing another two parameters $\delta_{\eta}$ and $\varepsilon_{\eta}$, we find $V_\eta$ using $O(\log(1/\varepsilon_\eta)/\delta_\eta)$ queries to $U_\eta$ that prepares $\eta'$ as a $(4\delta_{\eta}^{-1/2}, c, \Theta(\delta_{\eta}^{1/2}+\varepsilon_{\eta} \delta_{\eta}^{-1/2}))$-block-encoding of $\sqrt{\prescript{}{a}{\bra 0} \eta \ket 0_a}$, where $c = O(n+n_\rho+n_\sigma)$. Intuitively, $\sqrt{\prescript{}{a}{\bra 0} \eta \ket 0_a}$ is approximately proportional to $\sqrt{\sqrt{\sigma}\rho\sqrt{\sigma}}$ with a scaling factor $16\delta_{\sigma}^{-1}\delta_{\eta}^{-1}$.

\textbf{Step 4}. Estimate $\tr\left(\prescript{}{c}{\bra 0} \eta' \ket 0_c\right)$ through $V_\eta$ by quantum amplitude estimation \cite{BHMT02}. More precisely, we can obtain $\tilde x$ in $O(M)$ queries to $V_\eta$ (with high probability) such that
    \[
\abs{\tilde x - x} \leq \delta,
    \]
    where
\[
 \delta = 2 \pi \frac{\sqrt{x(1-x)}}{M} + \frac{\pi^2}{M^2}, ~\ x = \tr \left( \prescript{}{c}{\bra 0} \eta' \ket 0_c \right).
\]

\textbf{Step 5}.
Finally, we compute the value of $16 \tilde x / \sqrt{\delta_{\eta} \delta_{\sigma}}$ as our estimation of $F(\rho, \sigma)$. Intuitively, $\tilde x$ is an approximation of $\tr\left(\sqrt{\prescript{}{a}{\bra 0} \eta \ket 0_a}\right) \approx \sqrt{\delta_{\eta} \delta_{\sigma}} \tr\left(\sqrt{\sqrt{\sigma}\rho\sqrt{\sigma}}\right) / 16$ as mentioned in step 3.

\subsection{Error Analysis}

Now we are going to analyze the error of Algorithm \ref{algo}. Let $r = \min\{\rank(\rho), \rank(\sigma)\}$. First, we show that $\sigma'_b \rho \left(\sigma'_b\right)^\dag$ is a scaled approximation of $\sqrt{\sigma} \rho \sqrt{\sigma}$.

\begin{claim} \label{lemma:eta-approx}
    \[
    \Abs{16 \delta_\sigma^{-1} \sigma'_b \rho \left(\sigma'_b\right)^\dag - \sqrt{\sigma} \rho \sqrt{\sigma}} \leq \Theta(\delta_\sigma^{1/2}+\varepsilon_\sigma \delta_\sigma^{-1/2}).
    \]
\end{claim}
\begin{proof}
Note that
\begin{align*}
    16 \delta_\sigma^{-1} \sigma'_b \rho \left(\sigma'_b\right)^\dag - \sqrt{\sigma} \rho \sqrt{\sigma}
    =
     (4\delta_\sigma^{-1/2} \sigma'_b - \sqrt{\sigma}) \rho \left(4\delta_\sigma^{-1/2} \sigma'_b\right)^\dag
     +
    \sqrt{\sigma} \rho \left(\left(4\delta_\sigma^{-1/2} \sigma'_b\right)^\dag - \sqrt{\sigma}\right).
\end{align*}

By the triangle inequality for the operator norm that $\Abs{A + B} \leq \Abs{A} + \Abs{B}$ and the sub-multiplicativity that $\Abs{AB} \leq \Abs{A} \Abs{B}$, we have
\begin{align*}
    \Abs{16 \delta_\sigma^{-1} \sigma'_b \rho \left(\sigma'_b\right)^\dag - \sqrt{\sigma} \rho \sqrt{\sigma}}
    & \leq \Abs{ (4\delta_\sigma^{-1/2} \sigma'_b - \sqrt{\sigma}) \rho \left(4\delta_\sigma^{-1/2} \sigma'_b\right)^\dag }
    + \Abs{ \sqrt{\sigma} \rho \left(\left(4\delta_\sigma^{-1/2} \sigma'_b\right)^\dag - \sqrt{\sigma}\right) } \\
    & \leq \Abs{4\delta_\sigma^{-1/2} \sigma'_b - \sqrt{\sigma}} \Abs{\rho} \Abs{4\delta_\sigma^{-1/2} \sigma'_b} + \Abs{\sqrt{\sigma}} \Abs{\rho} \Abs{\left(4\delta_\sigma^{-1/2} \sigma'_b\right)^\dag - \sqrt{\sigma}}.
\end{align*}
Recall that $\sigma'_b = \prescript{}{b}{\bra 0} \sigma' \ket 0_b$, where $\sigma'$ is a $(4\delta_\sigma^{-1/2}, b, \Theta(\delta_\sigma^{1/2}+\varepsilon_\sigma \delta_\sigma^{-1/2}))$-block-encoding of $\sqrt{\sigma}$. That is,
\[
    \Abs{ 4\delta_\sigma^{-1/2} \sigma_b' - \sqrt{\sigma} } \leq \Theta\left(\delta_\sigma^{1/2}+\varepsilon_\sigma \delta_\sigma^{-1/2}\right).
\]
This gives that if $\delta_\sigma^{1/2}+\varepsilon_\sigma \delta_\sigma^{-1/2} < 1$, then we have
\[
    \Abs{4\delta_\sigma^{-1/2}\sigma_b'} \leq \Abs{4\delta_\sigma^{-1/2} \sigma_b' - \sqrt{\sigma}} + \Abs{\sqrt{\sigma}} \leq \Theta(1).
\]
Finally, together with $\Abs{\rho} \leq 1$, $\Abs{\sqrt{\sigma}} \leq 1$ and $\Abs{A} = \Abs{A^\dag}$, we have
\begin{align*}
    \Abs{16 \delta_\sigma^{-1} \sigma'_b \rho \left(\sigma'_b\right)^\dag - \sqrt{\sigma} \rho \sqrt{\sigma}}
    & \leq \Abs{4\delta_\sigma^{-1/2} \sigma_b' - \sqrt{\sigma}} \left( \Abs{4\delta_\sigma^{-1/2} \sigma_b'} + 1 \right) \\
    & \leq \Theta\left(\delta_\sigma^{1/2}+\varepsilon_\sigma \delta_\sigma^{-1/2}\right).
\end{align*}

\end{proof}

Next, we show how $\sqrt{\prescript{}{a}{\bra 0} \eta \ket 0_a}$ relates to the fidelity $F(\rho, \sigma)$.

\begin{claim} \label{lemma:eq-sigma}
\begin{align*}
    \abs{4\delta_\sigma^{-1/2} \tr \left( \sqrt{\prescript{}{a}{\bra 0} \eta \ket 0_a} \right) - F(\rho, \sigma)}
    \leq \Theta\left( r \sqrt{\delta_\sigma^{1/2}+\varepsilon_\sigma \delta_\sigma^{-1/2}} \right).
\end{align*}
\end{claim}
\begin{proof}
Let
\[
J = \frac{\delta_\sigma}{16} \sqrt\sigma\rho\sqrt\sigma - \prescript{}{a}{\bra 0} \eta \ket 0_a.
\]
In step 2 of the algorithm, it is shown in Claim \ref{lemma:eta-approx} that $\Abs{16 \delta_\sigma^{-1} \prescript{}{a}{\bra 0} \eta \ket 0_a - \sqrt{\sigma} \rho \sqrt{\sigma}} \leq \Theta(\delta_\sigma^{1/2}+\varepsilon_\sigma \delta_\sigma^{-1/2})$. This leads to
\[
\Abs{J} \leq \Theta(\delta_\sigma^{3/2}+\varepsilon_\sigma \delta_\sigma^{1/2}).
\]
We assume that the eigenvalues of ${\delta_\sigma}\sqrt\sigma\rho\sqrt\sigma/16$, $\prescript{}{a}{\bra 0} \eta \ket 0_a$ and $J$ are \begin{align*}&\mu_1 \geq \mu_2 \geq \dots \geq \mu_N,\\ &\nu_1 \geq \nu_2 \geq \dots \geq \nu_N,\\ &\xi_1 \geq \xi_2 \geq \dots \geq \xi_N,\end{align*} respectively. In our case, note that $\mu_{r + 1} = \dots = \mu_N = 0$ and $\nu_{r+1} = \dots = \nu_{N} = 0$. Since the three operators are all Hermitian, by Weyl's inequality, we have
\[
\nu_j + \xi_N \leq \mu_j \leq \nu_j + \xi_1
\]
for every $1 \leq j \leq N$. Now for each $j$, let us consider two cases:

{\vskip 3pt}

\textbf{Case 1}. $\nu_j \leq 2\Abs{J}$. In this case, $0 \leq \mu_j \leq 3\Abs{J}$, and then $\abs{\sqrt{\mu_j} - \sqrt{\nu_j}} \leq \sqrt{3\Abs{J}}$.

{\vskip 3pt}

\textbf{Case 2}. $\nu_j > 2\Abs{J}$. We have
\begin{align*}
\sqrt{\nu_j} - \sqrt{\Abs{J}} & \leq \sqrt{\nu_j - \Abs{J}} \leq \sqrt{\mu_j} \\
& \leq \sqrt{\nu_j + \Abs{J}} \leq \sqrt{\nu_j}+\sqrt{\Abs{J}}.
\end{align*}
Then it holds that $\abs{\sqrt{\mu_j} - \sqrt{\nu_j}} \leq \sqrt{\Abs{J}}$.

{\vskip 3pt}

The above two cases together yield that
\begin{align*}
    \abs{\tr\left(\sqrt{\prescript{}{a}{\bra 0} \eta \ket 0_a}\right) - \tr\left(\sqrt{\frac{\delta_\sigma}{16} \sqrt\sigma\rho\sqrt\sigma}\right)}
    = \abs{\sum_{j=1}^r \left(\sqrt{\mu_j} - \sqrt{\nu_j}\right)} \leq r \sqrt{3\Abs{J}}.
\end{align*}
These yield the proof.
\end{proof}

Finally, we establish the relationship between $\prescript{}{c}{\bra 0} \eta' \ket 0_c$ and $\sqrt{\prescript{}{a}{\bra 0} \eta \ket 0_a}$.

\begin{claim} \label{lemma:eq-eta}
\[
\abs{\tr \left( \prescript{}{c}{\bra 0} \eta' \ket 0_c \right) - \frac{\delta_{\eta}^{1/2}}{4} \tr \left( \sqrt{\prescript{}{a}{\bra 0} \eta \ket 0_a} \right)} \leq \Theta\left( r\left(\delta_\eta + \varepsilon_\eta\right) \right).
\]
\end{claim}
\begin{proof}
    In step 3 of the algorithm, we have $$\Abs{4 \delta_\eta^{-1/2} \prescript{}{c}{\bra 0} \eta' \ket 0_c - \sqrt{\prescript{}{a}{\bra 0} \eta \ket 0_a} } \leq \Theta(\delta_\eta^{1/2} + \varepsilon_\eta \delta_\eta^{-1/2}).$$ We note that $\prescript{}{a}{\bra 0} \eta \ket 0_a = \sigma'_b \rho \left(\sigma'_b\right)^\dag$, and thus $\rank\left(\sqrt{\prescript{}{a}{\bra 0} \eta \ket 0_a}\right) = \rank(\prescript{}{a}{\bra 0} \eta \ket 0_a) \leq \rank(\rho) \leq r$. For the same reason, we have $\rank(\prescript{}{c}{\bra 0} \eta' \ket 0_c) \leq r$. Therefore, we have
    \begin{align*}
    \abs{ 4 \delta_\eta^{-1/2} \tr\left( \prescript{}{c}{\bra 0} \eta' \ket 0_c\right) - \tr\left(\sqrt{\prescript{}{a}{\bra 0} \eta \ket 0_a}\right) }
    \leq \Theta(r(\delta_\eta^{1/2} + \varepsilon_\eta \delta_\eta^{-1/2})).
    \end{align*}
    These yield the proof.
\end{proof}

Combining the result of quantum amplitude estimation in step 4 of the algorithm with Claim \ref{lemma:eta-approx}, Claim \ref{lemma:eq-sigma} and Claim \ref{lemma:eq-eta}, we obtain an upper bound of the error of our estimation, which is
\begin{equation} \label{eq:error}
\begin{aligned}
    \abs{ \frac{16 \tilde x}{\sqrt{\delta_\eta\delta_\sigma}} - F(\rho, \sigma) }
    \leq
    \Theta \left( \frac{r(\delta_\eta+\varepsilon_\eta)+\delta} {\sqrt{\delta_\eta\delta_\sigma}} + r \sqrt{ \sqrt{\delta_\sigma} + \frac{\varepsilon_\sigma} {\sqrt{\delta_\sigma}} } \right).
\end{aligned}
\end{equation}

\subsection{Complexity}

In Algorithm \ref{algo}, the number of queries to $O_\rho$ and $O_\sigma$ is bounded by
\[
    O\left( \frac{1}{\delta_\sigma} \log \left(\frac{1}{\varepsilon_\sigma} \right) \cdot \frac{1}{\delta_\eta} \log \left(\frac{1}{\varepsilon_\eta} \right) \cdot M \right) = \tilde O_{\varepsilon_\sigma, \varepsilon_{\eta}}\left(\frac{M}{\delta_\sigma \delta_\eta}\right).
\]
In order to make the right hand side of Equation (\ref{eq:error}) $\leq \varepsilon$, we take $\delta_\sigma = \tilde \Theta(\varepsilon^4 / r^4)$, $\delta_\eta = \tilde \Theta(\varepsilon^6 / r^6)$, $\delta = \tilde \Theta(\varepsilon^6 / r^5)$, $\varepsilon_\sigma = \tilde \Theta(\varepsilon^4 / r^4)$, $\varepsilon_\eta = \tilde \Theta(\varepsilon^6 / r^6)$ and $M = \tilde \Theta(r^{2.5} / \varepsilon^{3.5})$ to minimize the number of queries
\[
    \tilde O_{\varepsilon_\sigma, \varepsilon_{\eta}}\left(\frac{M}{\delta_\sigma \delta_\eta}\right) = \tilde O_{r, \frac 1 \varepsilon}\left( \frac{r^{12.5}}{\varepsilon^{13.5}} \right).
\]
In addition, the number of elementary quantum gates is $\tilde O_{N, r, 1/ \varepsilon}(r^{12.5}/\varepsilon^{13.5}) = \poly(\log(N), r, 1/\varepsilon)$.

It can be seen that our algorithm exponentially outperforms the best known classical and even quantum algorithms for quantum fidelity estimation when $r$ is small, e.g., $r = \polylog(N)$. In spite of its large exponents of $r$ and $\varepsilon$ in the complexity, we believe that our algorithm can be applied on real-world problems, as several quantum-inspired algorithms proposed recently \cite{GLT18, Tan19} also with large exponents in their complexities are later shown to perform well in practice \cite{ADBL20}.

\section{Hardness} \label{sec:hardness}

Even though some quantum-inspired algorithms \cite{GLT18, Tan19} suggest that quantum exponential speedup can disappear in low-rank cases, quantum fidelity estimation still remains hard even under the low-rank assumption as discussed above. To show this, we first formally define \textsc{Low-Rank Fidelity Estimation} in the following.

\begin{problem} [\textsc{Low-Rank Fidelity Estimation}]
Given the description of two quantum circuits $O_\rho$ and $O_\sigma$ of size $\poly(n)$ that prepare purifications of $n$-qubit (mixed) quantum states $\rho$ and $\sigma$, respectively, where the rank of $\rho$ is $\poly(n)$, and the additive error $\varepsilon = 1/\poly(n)$, find an estimation of $F(\rho, \sigma)$ within additive error $\varepsilon$.
\end{problem}
Indeed, the description of quantum circuits is not required in our algorithm, but is needed for classical algorithms.
It was proved in \cite{CPCC20} that a variant of \textsc{Low-Rank Fidelity Estimation} is \textbf{DQC1}-hard, but the same proof also yields the \textbf{DQC1}-hardness of the problem stated here. It is known that \textbf{DQC1}-complete problems cannot be efficiently solved by classical computers unless the polynomial hierarchy collapses to the second level \cite{Mor17, FKM18}, which is commonly believed to be false. Therefore, our algorithm could be a candidate that shows the advantage of quantum computers over classical counterparts.

\section{Conclusion} \label{sec:conclusion}

In this paper, we proposed a quantum algorithm for quantum fidelity estimation, which yields an exponential speedup over the best known algorithms in the low-rank case. We hope it could be used as a subroutine in developing fidelity-based quantum algorithms \cite{SR21}. The exponents of some complexity factors in our algorithm are large, but we believe they could be reduced by some more sophisticated techniques (see, for example, \cite{Amb12}).
Furthermore, an interesting  problem is whether it is possible to
keep the advantage of exponential speedup in our algorithm with restricted quantum operations (e.g., Pauli measurements).

One of our main technical results (Theorem \ref{thm:sqrt}) can be extended to positive powers (not only square root) of positive semidefinite operator $A$, and therefore can be used in solving other problems, e.g., computing the sandwiched quantum R\'{e}nyi relative entropy \cite{WWY14, MDS13} for $0 < \alpha < 1$:
\[
\exp\left((\alpha - 1) D_\alpha(\rho \| \sigma)\right) = \tr \left( \left( \sigma^{\frac{1-\alpha}{2\alpha}} \rho  \sigma^{\frac{1-\alpha}{2\alpha}} \right)^{\alpha} \right),
\]
which reduces to the quantum state fidelity $F(\rho, \sigma)$ when $\alpha = 1/2$.

For the topics of future research, it would be interesting to try to adapt our quantum algorithms to computing other quantum information quantities with a similar form to the fidelity $F(\rho, \sigma) = \tr\left(\sqrt{\sqrt{\sigma}\rho\sqrt{\sigma}}\right)$, such as the von Neumann entropy $S(\rho) = -\tr(\rho \log \rho)$, the quantum relative von Neumann entropy $D(\rho\|\sigma) = \tr\left(\rho ( \log \rho - \log \sigma) \right)$, and the quantum relative min-entropy $-\log\left(\tr(\Pi_\rho \sigma)\right)$ \cite{Dat09}, where $\Pi_\rho$ is the projector onto the support of $\rho$.

\section*{Acknowledgment}
The authors would like to thank Professor Mark M. Wilde for pointing out the \textbf{QSZK}-hardness of fidelity estimation, and Minbo Gao for helpful discussions.

\bibliographystyle{unsrt}
\bibliography{arxiv}

\begin{thebibliography}{10}

\bibitem{Sho94}
P.~W. Shor.
\newblock Algorithms for quantum computation: discrete logarithms and
  factoring.
\newblock In {\em Proceedings of the 35th Annual Symposium on Foundations of
  Computer Science}, pages 124--134, 1994.

\bibitem{HHL09}
A.~W. Harrow, A.~Hassidim, and S.~Lloyd.
\newblock Quantum algorithm for linear systems of equations.
\newblock {\em Physical Review Letters}, 103(15):150502, 2009.

\bibitem{Fey82}
R.~P. Feynman.
\newblock Simulating physics with computers.
\newblock {\em International Journal of Theoretical Physics}, 21:467--488,
  1982.

\bibitem{AL97}
D.~S. Abrams and S.~Lloyd.
\newblock Simulation of many-body fermi systems on a universal quantum
  computer.
\newblock {\em Physical Review Letters}, 79(13):2586--2589, 1997.

\bibitem{FKW02}
M.~Freedman, A.~Kitaev, and Z.~Wang.
\newblock Simulation of topological field theories by quantum computers.
\newblock {\em Communications in Mathematical Physics}, 227(3):587--603, 2002.

\bibitem{KJL08}
I.~Kassal, S.~P. Jordan, P.~J. Love, M.~Mohseni, and A~Aspuru-Guzik.
\newblock Polynomial-time quantum algorithm for the simulation of chemical
  dynamics.
\newblock {\em Proceedings of the National Academy of Sciences of the United
  States of America}, 105(48):18681--86, 2008.

\bibitem{Uhl76}
A.~Uhlmann.
\newblock The ``transition probability'' in the state space of a {*}-algebra.
\newblock {\em Reports on Mathematical Physics}, 9(2):273--279, 1976.

\bibitem{Joz94}
R.~Jozsa.
\newblock Fidelity for mixed quantum states.
\newblock {\em Journal of Modern Optics}, 41(12):2315--2323, 1994.

\bibitem{NC10}
M.~A. Nielsen and I.~L. Chuang.
\newblock {\em Quantum Computation and Quantum Information}.
\newblock Cambridge University Press, 2010.

\bibitem{TV11}
B.~T. Torosov and N.~V. Vitanov.
\newblock Smooth composite pulses for high-fidelity quantum information
  processing.
\newblock {\em Physical Review A}, 83(5):053420, 2011.

\bibitem{RCR21}
T.~F. Roque, A.~A. Clerk, and H.~Ribeiro.
\newblock Engineering fast high-fidelity quantum operations with constrained
  interactions.
\newblock {\em npj Quantum Information}, 7(1):1--17, 2021.

\bibitem{SR21}
F.~Shahi and A.~T. Rezakhani.
\newblock Fidelity-based supervised and unsupervised learning for binary
  classification of quantum states.
\newblock {\em The European Physical Journal Plus}, 136:280, 2021.

\bibitem{Wat02}
J.~Watrous.
\newblock Limits on the power of quantum statistical zero-knowledge.
\newblock In {\em Proceedings of the 43rd Annual IEEE Symposium on Foundations
  of Computer Science}, pages 459--468, 2002.

\bibitem{BKL+19}
F.~G. S.~L. Brand\~{a}o, A.~Kalev, T.~Li, C.~Y.-Y. Lin, K.~M. Svore, and X.~Wu.
\newblock Quantum {SDP} solvers: Large speed-ups, optimality, and applications
  to quantum learning.
\newblock In {\em Proceedings of the 46th International Colloquium on Automata,
  Languages, and Programming}, pages 27:1--27:14, 2019.

\bibitem{vAG19}
J.~van Apeldoorn and A.~Gily\'{e}n.
\newblock Improvements in quantum {SDP}-solving with applications.
\newblock In {\em Proceedings of the 46th International Colloquium on Automata,
  Languages, and Programming}, pages 99:1--99:15, 2019.

\bibitem{GL20}
A.~{Gily\'{e}n} and T.~Li.
\newblock Distributional property testing in a quantum world.
\newblock In {\em Proceedings of the 11th Innovations in Theoretical Computer
  Science Conference}, volume 151, pages 25:1--25:19, 2020.

\bibitem{GLM+20}
A.~Gily\'{e}n, S.~Lloyd, I.~Marvian, Y.~Quek, and M.~M. Wilde.
\newblock Quantum algorithm for {Petz} recovery channels and pretty good
  measurements.
\newblock {\em Physical Review Letters}, 128(22):220502, 2022.

\bibitem{ARSW21}
R.~Agarwal, S.~Rethinasamy, K.~Sharma, and M.~M. Wilde.
\newblock Estimating distinguishability measures on quantum computers.
\newblock ArXiv e-prints, 2021.
\newblock arXiv:2108.08406.

\bibitem{GHS21}
T.~Gur, M.~Hsieh, and S.~Subramanian.
\newblock Sublinear quantum algorithms for estimating von {Neumann} entropy.
\newblock ArXiv e-prints, 2021.
\newblock arXiv:2111.11139.

\bibitem{SH21}
S.~Subramanian and M.~Hsieh.
\newblock Quantum algorithm for estimating $\alpha$-{Renyi} entropies of
  quantum states.
\newblock {\em Physical Review A}, 104(2):022428, 2021.

\bibitem{Wat08}
J.~Watrous.
\newblock Quantum computational complexity.
\newblock ArXiv e-prints, 2008.
\newblock arXiv:0804.3401.

\bibitem{DM97}
V.V. Dodonov and V.I. Man'ko.
\newblock Positive distribution description for spin states.
\newblock {\em Physics Letters A}, 229(6):335--339, 1997.

\bibitem{dAMP03}
G.~M. d’Ariano, L.~Maccone, and M.~Paini.
\newblock Spin tomography.
\newblock {\em Journal of Optics B: Quantum and Semiclassical Optics}, 5(1):77,
  2003.

\bibitem{GLF10}
D.~Gross, Y.~K. Liu, S.~T. Flammia, S.~Becker, and J.~Eisert.
\newblock Quantum state tomography via compressed sensing.
\newblock {\em Physical Review Letters}, 105(15):150401, 2010.

\bibitem{OW16}
R.~O'Donnell and J.~Wright.
\newblock Efficient quantum tomography.
\newblock In {\em Proceedings of the 48th ACM Symposium on Theory of
  Computing}, pages 899--912, 2016.

\bibitem{HHJ+17}
J.~Haah, A.~W. Harrow, Z.~Ji, X.~Wu, and N.~Yu.
\newblock Sample-optimal tomography of quantum states.
\newblock {\em IEEE Transactions on Information Theory}, 63(9):5628--5641,
  2017.

\bibitem{BCW01}
H.~Buhrman, R.~Cleve, J.~Watrous, and R.~de~Wolf.
\newblock Quantum fingerprinting.
\newblock {\em Physical Review Letters}, 87(16):167902, 2001.

\bibitem{KHC19}
M.~Kang, J.~Heo, S.~Choi, S.~Moon, and S.~Han.
\newblock Implementation of {SWAP} test for two unknown states in photons via
  cross-{K}err nonlinearities under decoherence effect.
\newblock {\em Scientific Reports}, 9(1):6167, 2019.

\bibitem{TYKI06}
Y.~Tokunaga, T.~Yamamoto, M.~Koashi, and N.~Imoto.
\newblock Fidelity estimation and entanglement verification for experimentally
  produced four-qubit cluster states.
\newblock {\em Physical Review A}, 74:020301(R), 2006.

\bibitem{GLGP07}
O.~G\"{u}hne, C.-Y. Lu, W.-B. Gao, and J.-W. Pan.
\newblock Toolbox for entanglement detection and fidelity estimation.
\newblock {\em Physical Review A}, 76:030305(R), 2007.

\bibitem{GT09}
O.~G\"{u}hne and G.~T\'{o}th.
\newblock Entanglement detection.
\newblock {\em Physics Reports}, 474(1-6):1--75, 2009.

\bibitem{FL11}
S.~T. Flammia and Y.-K. Liu.
\newblock Direct fidelity estimation from few pauli measurements.
\newblock {\em Physical Review Letters}, 106:230501, 2011.

\bibitem{dSLP11}
M.~P. da~Silva, O.~Landon-Cardinal, and D.~Poulin.
\newblock Practical characterization of quantum devices without tomography.
\newblock {\em Physical Review Letters}, 107:210404, 2011.

\bibitem{CPCC20}
M.~Cerezo, A.~Poremba, L.~Cincio, and P.~J. Coles.
\newblock Variational quantum fidelity estimation.
\newblock {\em Quantum}, 4:248, 2020.

\bibitem{CSZW22}
R.~Chen, Z.~Song, X.~Zhao, and X.~Wang.
\newblock Variational quantum algorithms for trace distance and fidelity
  estimation.
\newblock {\em Quantum Science and Technology}, 7(1):015019, 2022.

\bibitem{TV21}
K.~C. Tan and T.~Volkoff.
\newblock Variational quantum algorithms to estimate rank, quantum entropies,
  fidelity, and fisher information via purity minimization.
\newblock {\em Physical Review Research}, 3(3):033251, 2021.

\bibitem{BK21}
L.~Bittel and M.~Kliesch.
\newblock Training variational quantum algorithms is {NP}-hard.
\newblock {\em Physical Review Letters}, 127(12):120502, 2021.

\bibitem{KL98}
E.~Knill and R.~Laflamme.
\newblock Power of one bit of quantum information.
\newblock {\em Physical Review Letters}, 81:5672, 1998.

\bibitem{Mor17}
T.~Morimae.
\newblock Hardness of classically sampling the one-clean-qubit model with
  constant total variation distance error.
\newblock {\em Physical Review A}, 96:040302(R), 2017.

\bibitem{FKM18}
K.~Fujii, H.~Kobayashi, T.~Morimae, H.~Nishimura, S.~Tamate, and S.~Tani.
\newblock Impossibility of classically simulating one-clean-qubit model with
  multiplicative error.
\newblock {\em Physical Review Letters}, 120:200502, 2018.

\bibitem{LC19}
G.~H. Low and I.~L. Chuang.
\newblock Hamiltonian simulation by qubitization.
\newblock {\em Quantum}, 3:163, 2019.

\bibitem{GSLW19}
A.~Gily\'{e}n, Y.~Su, G.~H. Low, and N.~Wiebe.
\newblock Quantum singular value transformation and beyond: exponential
  improvements for quantum matrix arithmetics.
\newblock In {\em Proceedings of the 51st Annual ACM SIGACT Symposium on Theory
  of Computing}, pages 193--204, 2019.

\bibitem{BHMT02}
G.~Brassard, P~H{\o}yer, M.~Mosca, and A.~Tapp.
\newblock Quantum amplitude amplification and estimation.
\newblock {\em Quantum Computation and Information}, 305:53--74, 2002.

\bibitem{LH08}
H.~Li and F.~D.~M. Haldane.
\newblock Entanglement spectrum as a generalization of entanglement entropy:
  Identification of topological order in non-{A}belian fractional quantum hall
  effect states.
\newblock {\em Physical Review Letters}, 101(1):010504, 2008.

\bibitem{MdW16}
A.~Montanaro and R.~de~Wolf.
\newblock A survey of quantum property testing.
\newblock {\em Theory of Computing Library, Graduate Surveys}, 7:1--81, 2016.

\bibitem{BOW19}
C.~B{\v{a}}descu, R.~O'Donnell, and J.~Wright.
\newblock Quantum state certification.
\newblock In {\em Proceedings of the 51st ACM Symposium on Theory of
  Computing}, pages 503--514, 2019.

\bibitem{OW15}
R.~O'Donnell and J.~Wright.
\newblock Quantum spectrum testing.
\newblock In {\em Proceedings of the 47th ACM Symposium on Theory of
  Computing}, pages 529--538, 2015.

\bibitem{OW17}
R.~O'Donnell and J.~Wright.
\newblock Efficient quantum tomography {II}.
\newblock In {\em Proceedings of the 49th ACM Symposium on Theory of
  Computing}, pages 962--974, 2017.

\bibitem{KLL+17}
S.~Kimmel, C.~Y. Lin, G.~H. Low, M.~Ozols, and T.~J. Yoder.
\newblock Hamiltonian simulation with optimal sample complexity.
\newblock {\em npj Quantum Information}, 3(1):1--7, 2017.

\bibitem{AISW19}
J.~Acharya, I.~Issa, N.~V. Shende, and A.~B. Wagne.
\newblock Measuring quantum entropy.
\newblock In {\em 2019 IEEE International Symposium on Information Theory},
  pages 3012--3016, 2019.

\bibitem{BHH11}
S.~Bravyi, A.~W. Harrow, and A.~Hassidim.
\newblock Quantum algorithms for testing properties of distributions.
\newblock {\em IEEE Transactions on Information Theory}, 57(6):3971--3981,
  2011.

\bibitem{Mon15}
A.~Montanaro.
\newblock Quantum speedup of {Monte} {Carlo} methods.
\newblock {\em Proceedings of the Royal Society A}, 471(2181):20150301, 2015.

\bibitem{CFMdW10}
S.~Chakraborty, E.~Fischer, A.~Matsliah, and R.~de~Wolf.
\newblock New results on quantum property testing.
\newblock In {\em Proceedings of the 30th International Conference on
  Foundations of Software Technology and Theoretical Computer Science},
  volume~8, pages 145--156, 2010.

\bibitem{LW19}
T.~Li and X.~Wu.
\newblock Quantum query complexity of entropy estimation.
\newblock {\em IEEE Transactions on Information Theory}, 65(5):2899--2921,
  2019.

\bibitem{WGL+22}
Q.~Wang, J.~Guan, J.~Liu, Z.~Zhang, and M.~Ying.
\newblock New quantum algorithms for computing quantum entropies and distances.
\newblock ArXiv e-prints, 2022.
\newblock arXiv:2203.13522.

\bibitem{GP22}
A.~{Gily\'{e}n} and A.~Poremba.
\newblock Improved quantum algorithms for fidelity estimation.
\newblock ArXiv e-prints, 2022.
\newblock arXiv:2203.15993.

\bibitem{LMR14}
S.~Lloyd, M.~Mohseni, and P.~Rebentrost.
\newblock Quantum principal component analysis.
\newblock {\em Nature Physics}, 10:631--633, 2014.

\bibitem{CGJ19}
S.~Chakraborty, A.~Gily\'{e}n, and S.~Jeffery.
\newblock The power of block-encoded matrix powers: improved regression
  techniques via faster hamiltonian simulation.
\newblock In {\em Proceedings of the 46th International Colloquium on Automata,
  Languages, and Programming}, pages 33:1--33:14, 2019.

\bibitem{Gil19}
A.~Gily\'{e}n.
\newblock {\em Quantum Singular Value Transformation \& Its Algorithmic
  Applications}.
\newblock PhD thesis, University of Amsterdam, Amsterdam: Institute for Logic,
  Language and Computation, 2019.

\bibitem{GLT18}
A.~Gily\'{e}n, S.~Lloyd, and E.~Tang.
\newblock Quantum-inspired low-rank stochastic regression with logarithmic
  dependence on the dimension.
\newblock ArXiv e-prints, 2018.
\newblock arXiv:1811.04909.

\bibitem{Tan19}
E.~Tang.
\newblock A quantum-inspired classical algorithm for recommendation systems.
\newblock In {\em Proceedings of the 51st Annual Symposium on Theory of
  Computing}, pages 219--228, 2019.

\bibitem{ADBL20}
J.~M. Arrazola, A.~Delgado, B.~R. Bardhan, and S.~Lloyd.
\newblock Quantum-inspired algorithms in practice.
\newblock {\em Quantum}, 4:307, 2020.

\bibitem{Amb12}
A.~Ambainis.
\newblock Variable time amplitude amplification and quantum algorithms for
  linear algebra problems.
\newblock In {\em Proceedings of the 29th Symposium on Theoretical Aspects of
  Computer Science}, pages 636--647, 2012.

\bibitem{WWY14}
M.~M. Wilde, A.~Winter, and D.~Yang.
\newblock Strong converse for the classical capacity of entanglement-breaking
  and {H}adamard channels via a sandwiched {R}\'{e}nyi relative entropy.
\newblock {\em Communications in Mathematical Physics}, 331(2):593--622, 2014.

\bibitem{MDS13}
M.~M\"{u}ller-Lennert, F.~Dupuis, O.~Szehr, S.~Fehr, and M.~Tomamichel.
\newblock On quantum {R}\'{e}nyi entropies: A new generalization and some
  properties.
\newblock {\em Journal of Mathematical Physics}, 54(12):122203, 2013.

\bibitem{Dat09}
N.~Datta.
\newblock Min- and max-relative entropies and a new entanglement monotone.
\newblock {\em IEEE Transactions on Information Theory}, 55(6):2816--2826,
  2009.

\end{thebibliography}

\end{document}